\documentclass[12pt]{article}

\usepackage{dina4pcg}
\usepackage{epsfig}
\usepackage{amssymb}

\def\oas{{\cal O}(\as)}
\def\yc{y_{\rm cut}}
\def\oass{{\cal O}(\as^2)}
\def\oo{\'o}

\def\mz{M_Z}
\def\asz{\as(\mz)}
\def\as{\alpha_s}
\def\z0{Z^0}
\def\qq{q\bar q}
\def\q2{Q^2}

\def\etjet{E_T^{\rm jet}}
\def\etajet{\eta^{\rm jet}}
\def\kt{k_T}
\def\g2{GeV$^2$}

\def\etsbj{E_T^{\rm sbj}}
\def\etasbj{\eta^{\rm sbj}}
\def\phisbj{\phi^{\rm sbj}}
\def\phijet{\phi^{\rm jet}}
\def\asbj{\alpha^{\rm sbj}}
\def\colab#1{#1 Coll.}
\def\etal{et al.}

\def\Journal#1#2#3#4{{#1} {#2} (#3) #4}

\def\NPB{{\em Nucl. Phys.} {\bf B}}
\def\PLB{{\em Phys. Lett.}  {\bf B}}
\def\PRL{{\em Phys. Rev. Lett.}}
\def\PRD{{\em Phys. Rev.} {\bf D}}

\def\ZPC{{\em Z. Phys.} {\bf C}}

\def\EPC{{\em Eur. Phys. Jour.} {\bf C}}

\def\CPC{{\em Comp. Phys. Comm.}}

\def\JHEP{JHEP}

\begin{document}

\title{\bf Event shapes and subjet distributions at
  HERA\footnote{Talk given at HEP2005 International Europhysics
    Conference on High Energy Physics, Lisbon, Portugal, July $21^{\rm
      st} - 27^{\rm th}$, 2005.}}

\author{C. Glasman\thanks{Ram\oo n y Cajal Fellow.}\\
(on behalf of the ZEUS Collaboration)\\
Universidad Aut\oo noma de Madrid, Spain}

\date{}

\maketitle

\begin{abstract}
Recent results on subjet distributions from ZEUS are presented. The
measured normalised cross sections were used to study the pattern of
parton radiation. The comparison of the measurements with
leading-logarithm parton-shower Monte Carlo models and perturbative
QCD calculations shows a good agreement between data and
predictions. Results on event-shape mean values and distributions are
also presented. These measurements were used to test the predictions
of the power-correction model for hadronisation. A universal value,
within $10\%$, of the effective parameter $\bar\alpha_0$ of the model
was obtained.
\end{abstract}

\section{Introduction}
\begin{figure}[ht]
\setlength{\unitlength}{1.0cm}
\begin{picture} (10.0,5.5)
\put (1.0,1.0){\epsfig{figure=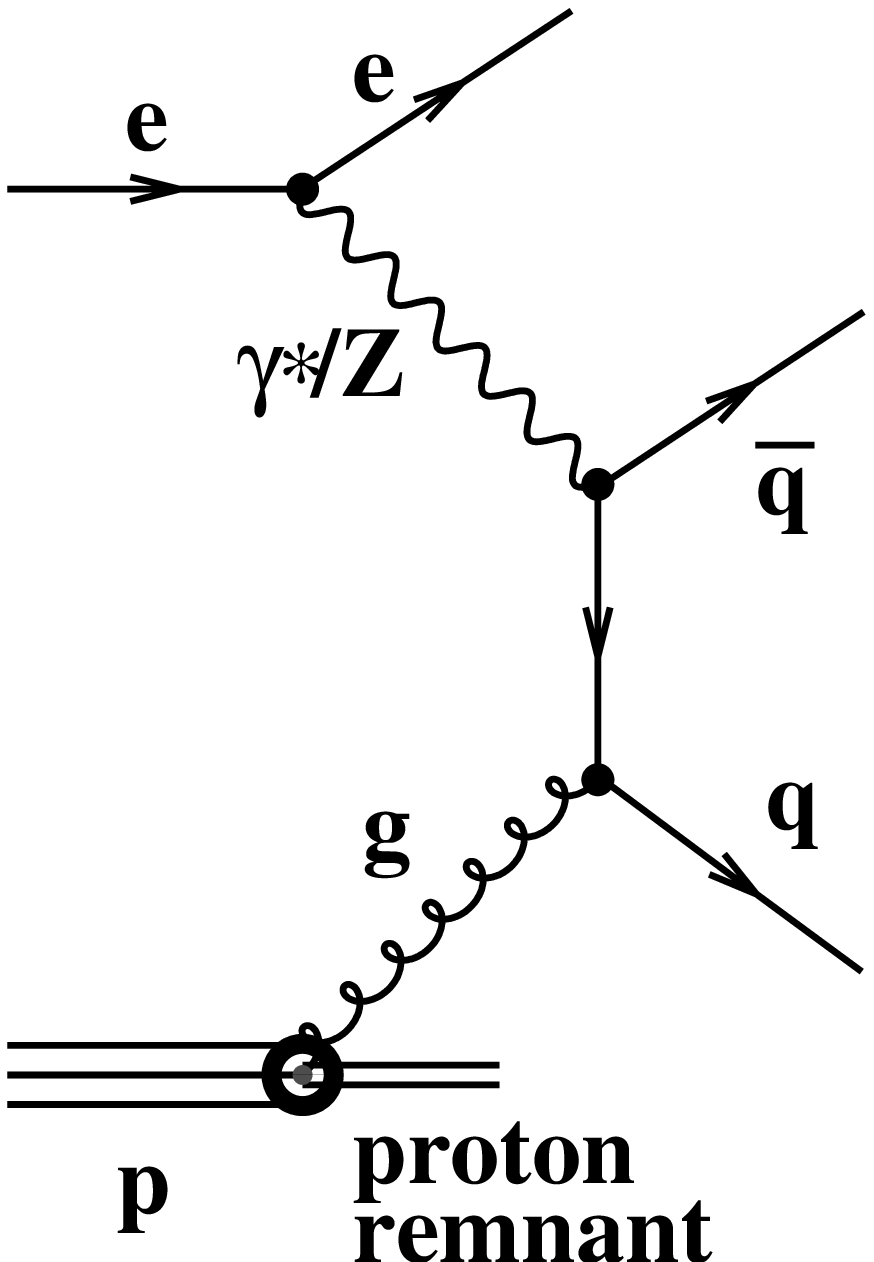,width=3.0cm}}
\put (5.5,1.0){\epsfig{figure=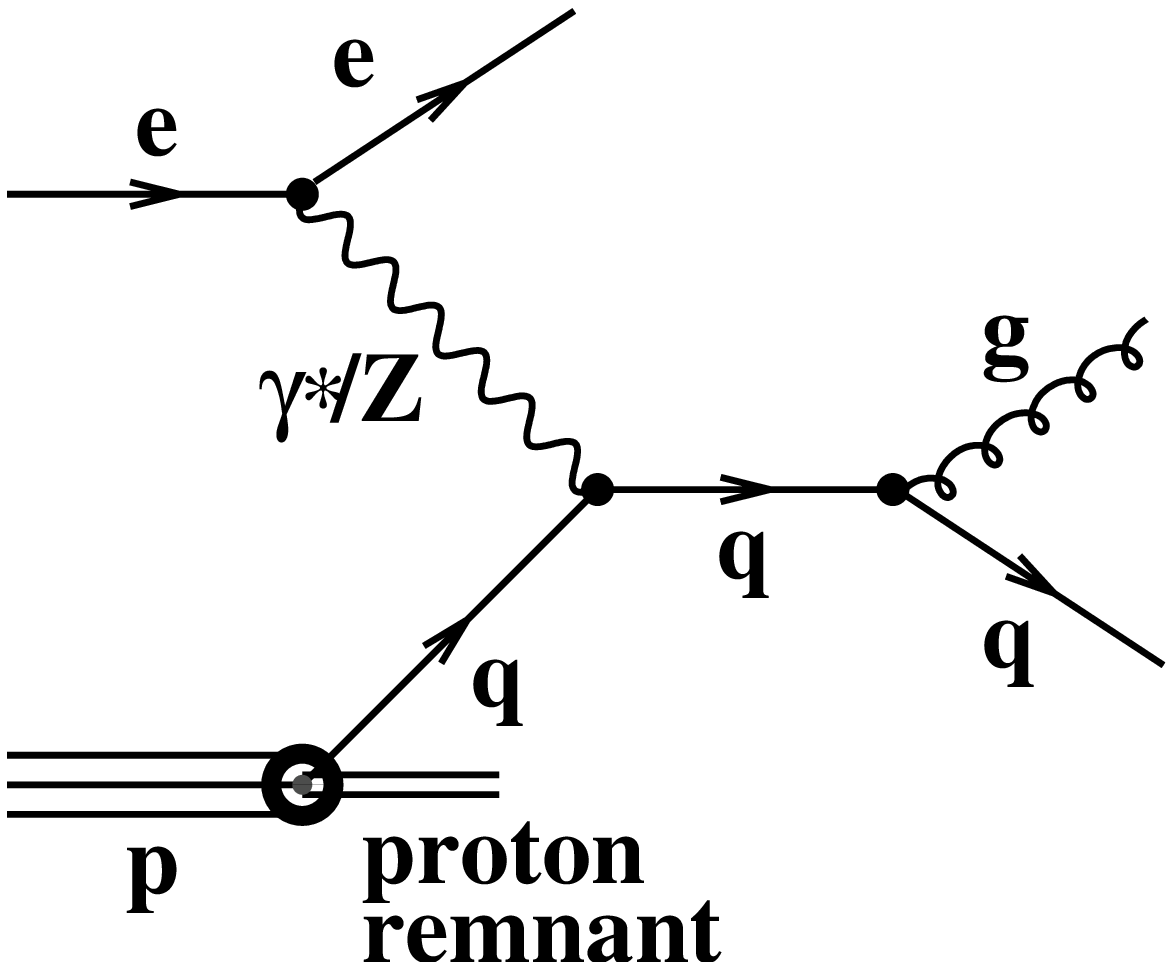,width=5.0cm}}
\put (11.0,1.0){\epsfig{figure=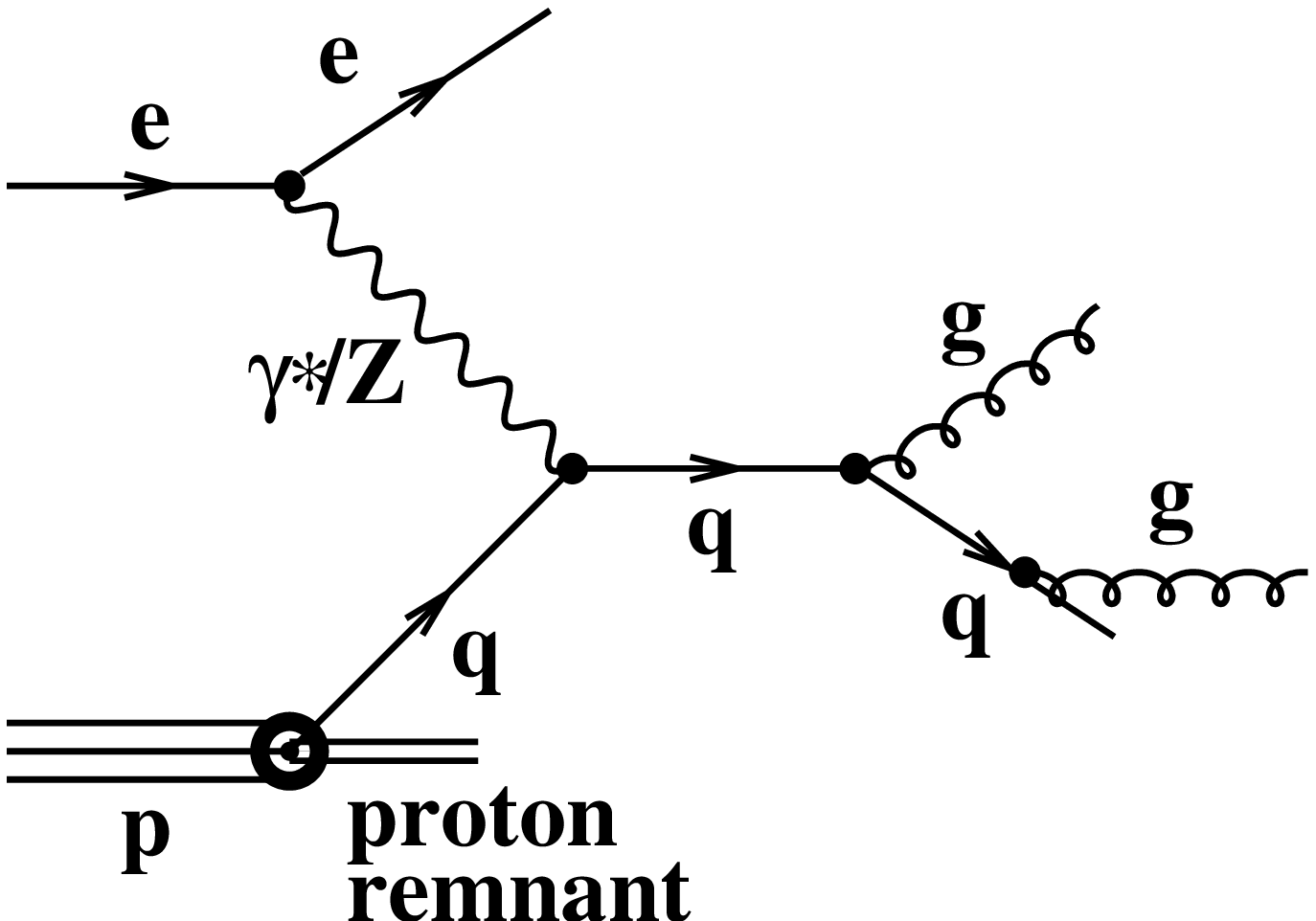,width=6.0cm}}
\put (2.0,0.0){\bf\small (a)}
\put (8.5,0.0){\bf\small (b)}
\put (14.0,0.0){\bf\small (c)}
\end{picture}
\caption{Examples of Feynman diagrams for deep inelastic $ep$
  scattering processes at lowest-order QCD: (a) boson-gluon fusion and
  (c) QCD Compton. (c) Example of a higher-order diagram.
  \label{one}}
\end{figure}

At lowest-order QCD, the diagrams that contribute to neutral current
(NC) deep inelastic $ep$ scattering (DIS) at HERA are the boson-gluon
fusion (BGF) ($Vg\rightarrow \qq$, where $V=\gamma^*$ or $\z0$,
Fig.~\ref{one}a) and QCD-Compton (QCDC) ($Vq\rightarrow qg$,
Fig.~\ref{one}b) processes. Figure~\ref{one}c shows an example of a
higher-order diagram. Any observable can be expressed as the
convolution of the parton densities in the proton, $f_a$, times the
matrix elements, 
$$\sum_a f_a(x,\mu_F)\otimes {\rm Matrix\ Elements},$$
where $\mu_F$ is the factorisation scale. The matrix elements, in
addition to $x$ and $\mu_F$, depend also on the strong coupling
constant, $\as$, and the renormalisation scale $\mu_R$. The sum runs
over all type of partons $a$. Thus, in the regions of phase space
where the parton densities are well
constrained, measurements of e.g. jet cross
sections can be used to perform tests of perturbative QCD (pQCD) and
determinations of $\as$.

The hadronic final state in NC DIS can also be used to study the
pattern of parton radiation by means e.g. of subjets inside
jets. Subjets observables are calculable in pQCD and so they provide
stringent tests of the theory. On the other hand, the hadronisation
process, a non-perturbative effect, can be studied by means of the
event shapes. Recent developments on the model of power-law
corrections~\cite{power} have prompted revived interest in
understanding hadronisation within the framework of pQCD.

\section{Subjet distributions}
The investigation of the internal structure of jets gives insight into
the transition between a parton produced in a hard process and the
experimentally observable jet of hadrons. At sufficiently high jet
transverse energy, $\etjet$, where the effects of fragmentation can be
neglected, the jet structure can be calculated perturbatively. The
lowest non-trivial-order contribution to the jet substructure is given
by $\oas$ calculations for NC DIS in the laboratory (LAB)
frame. Next-to-leading-order (NLO) calculations of jet substructure
can be obtained in the LAB frame since, in such a case, it is possible
to have three partons inside one jet.

The $\kt$ cluster algorithm~\cite{np:b406:187} was used in the
longitudinally invariant inclusive mode~\cite{pr:d48:3160} to define
jets in the hadronic final state. The internal structure of the jets
can be studied by means of the subjet topology. Subjets were resolved
within a jet by considering all particles associated with the jet and
repeating the application of the $\kt$ algorithm until, for every pair
of particles $i$ and $j$ the quantity
$$d_{ij}={\rm min}(E_{T,i},E_{T,j})^2\cdot[(\eta_i-\eta_j)^2+(\phi_i-\phi_j)^2],$$
where $E_{T,i}$, $\eta_i$ and $\phi_i$ are the transverse energy,
pseudorapidity and azimuth of particle $i$, respectively, was greater
than $d_{\rm cut}=\yc\cdot(\etjet)^2$. All remaining clusters were
called subjets. The subjet multiplicity depends upon the value chosen
for the resolution parameter $\yc$.

The pattern of QCD radiation from a primary parton has been
studied~\cite{subjets} by measuring normalised cross sections as a
function of subjet observables: the ratio between the subjet
transverse energy and that of the jet, $\etsbj/\etjet$, the difference
between the subjet pseudorapidity (azimuth) and that of the jet,
$\etasbj-\etajet$ ($|\phisbj-\phijet|$), and $\asbj$, the angle, as
viewed from the jet centre, between the highest transverse energy
subjet and the beam line in the pseudorapidity-azimuth plane. The
measurements were done for $\q2>125$ \g2, where $\q2$ is the momentum
transfer. Jets of $\etjet>14$ GeV and $-1<\etajet<2.5$ were
selected. The final sample consisted of those jets which had two
subjets for $\yc=0.05$.

The $\oas$ and $\oass$ QCD calculations used to compare with the data
are based on the program {\sc Disent}~\cite{disent}. For these
calculations, the number of flavours was set to five; the
renormalisation and factorisation scales were both set to
$\mu_R=\mu_F=Q$; $\as$ was calculated at two loops using  
$\Lambda^{(5)}_{\overline{\rm MS}}=220$~MeV, which corresponds to 
$\asz=0.1175$. The MRST99~\cite{epj:c4:463} parameterisations of the
proton parton density functions (PDFs) were used.

\begin{figure}[ht]
\setlength{\unitlength}{1.0cm}
\begin{picture} (10.0,10.0)
\put (1.0,0.5){\epsfig{figure=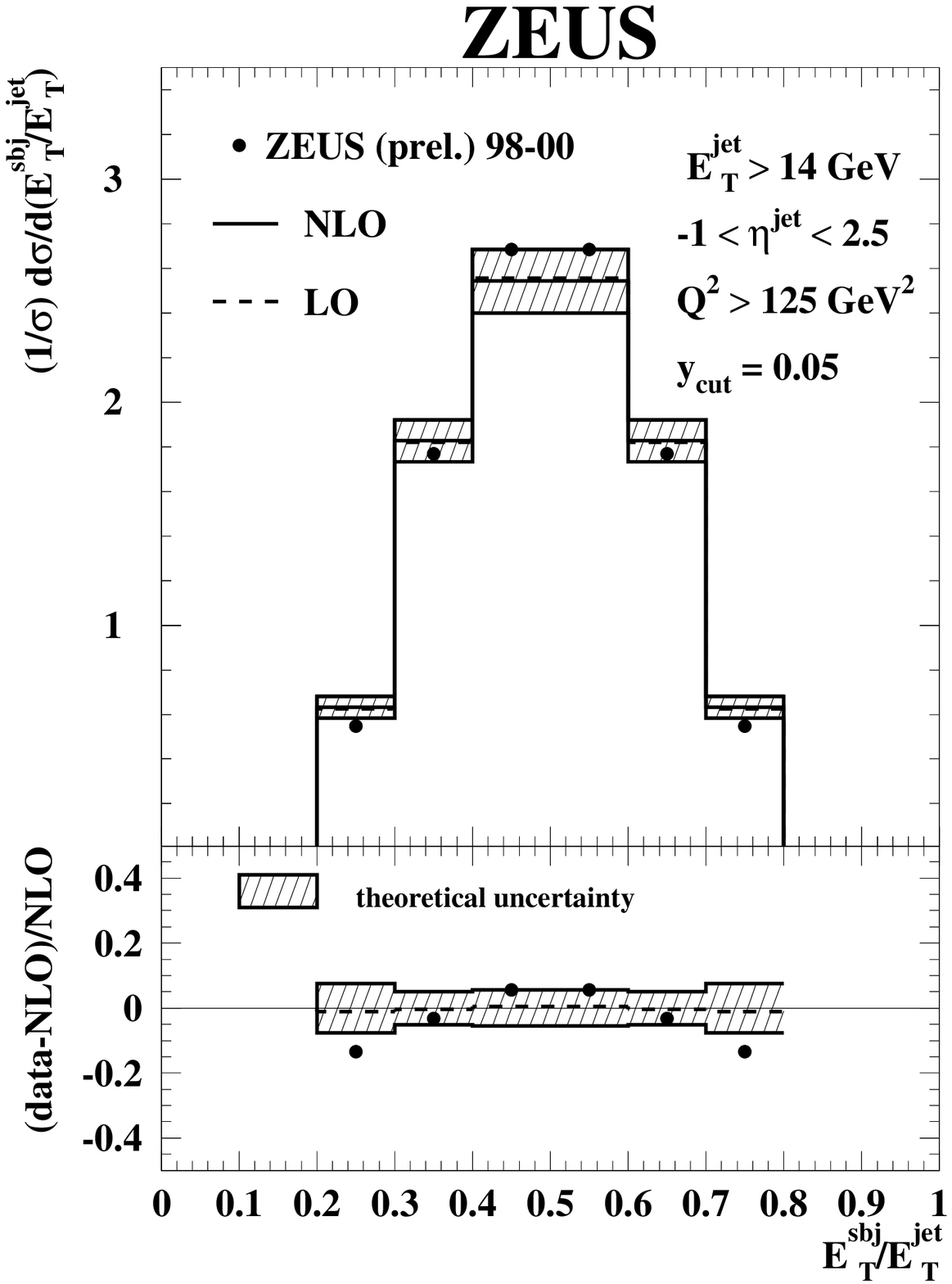,width=9.0cm}}
\put (9.0,0.5){\epsfig{figure=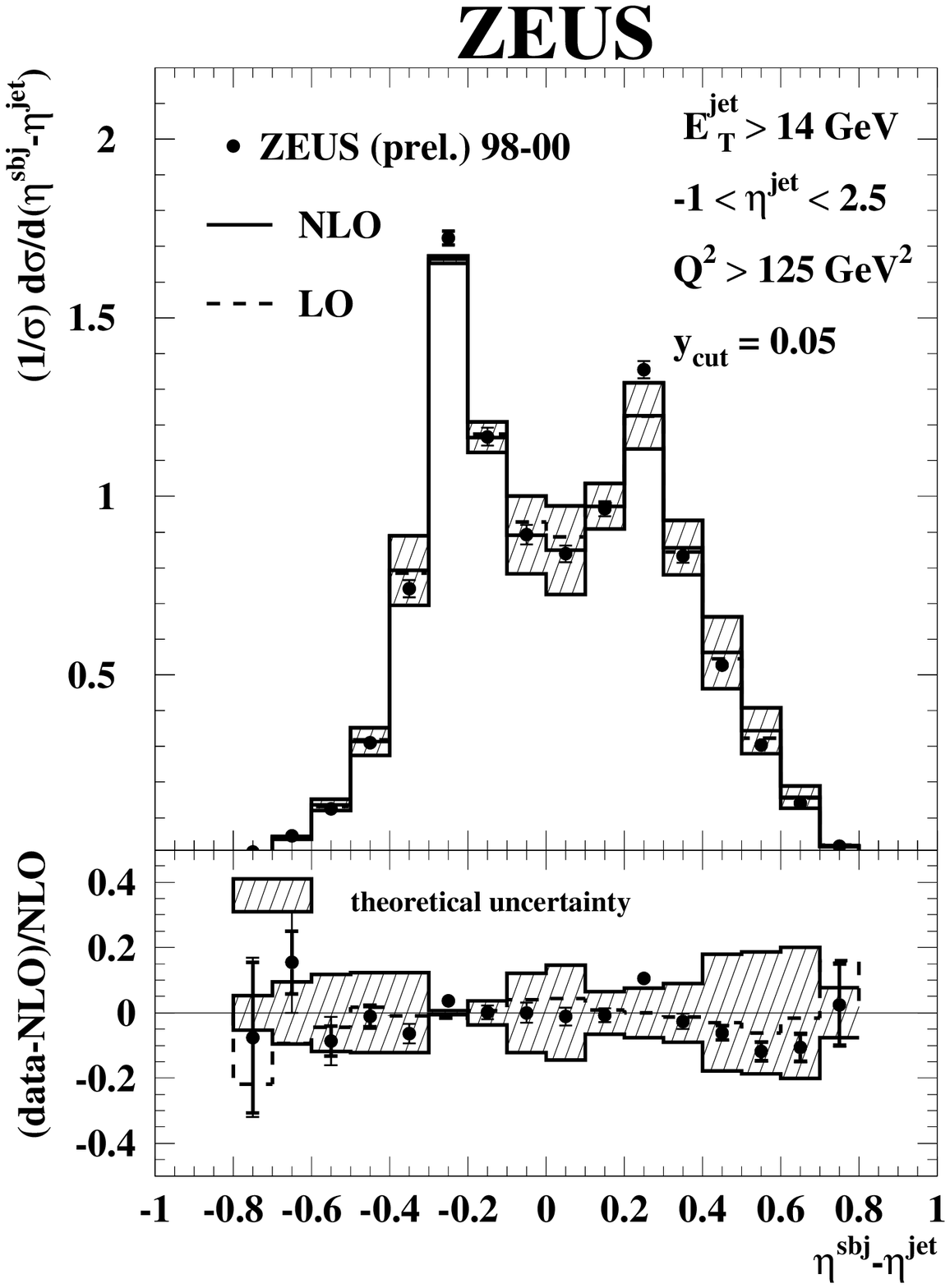,width=9.0cm}}
\put (5.25,0.0){\bf\small (a)}
\put (13.25,0.0){\bf\small (b)}
\end{picture}
\caption{Normalised differential subjet cross sections as functions of
  (a) $\etsbj/\etjet$ and (b) $\etasbj-\etajet$. For comparison, the
  predictions of {\sc Disent} at LO (dashed histograms) and NLO (solid
  histograms) are included.
  \label{two}}
\end{figure}

\begin{figure}[ht]
\setlength{\unitlength}{1.0cm}
\begin{picture} (10.0,9.5)
\put (1.0,0.5){\epsfig{figure=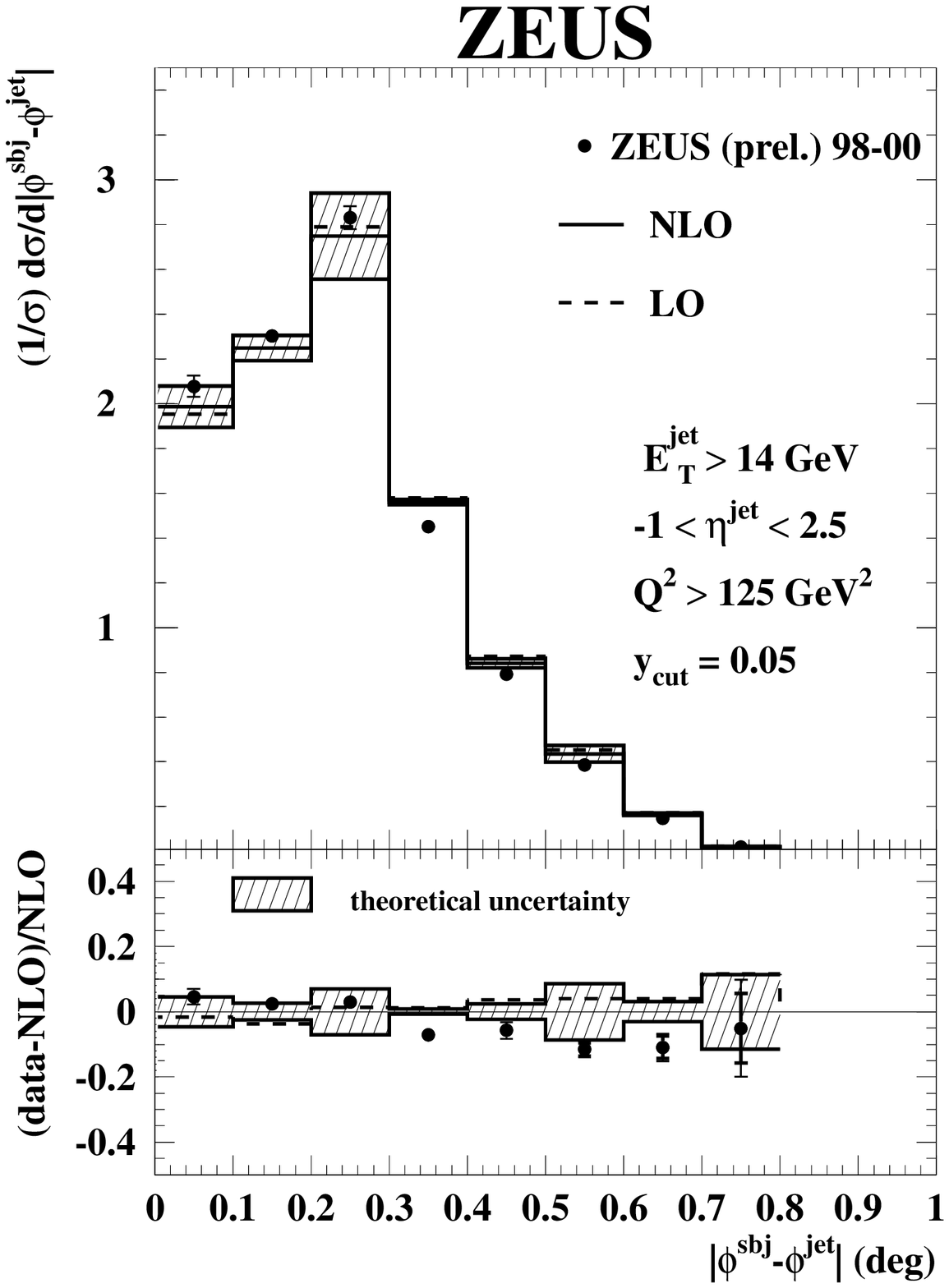,width=9.0cm}}
\put (9.0,0.5){\epsfig{figure=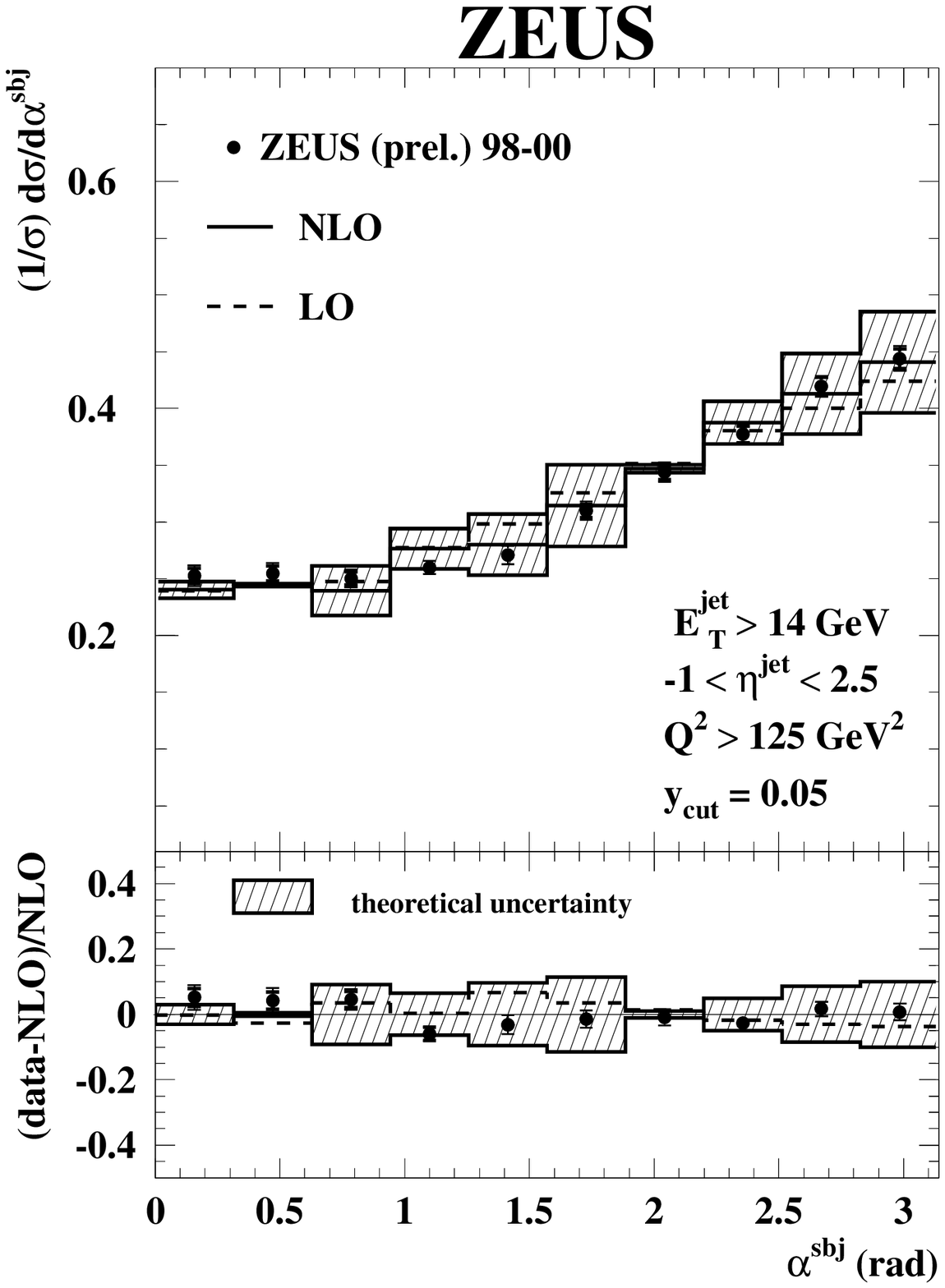,width=9.0cm}}
\put (5.25,0.0){\bf\small (a)}
\put (13.25,0.0){\bf\small (b)}
\end{picture}
\caption{Normalised differential subjet cross sections as functions of
  (a) $|\phisbj-\phijet|$ and (b) $\asbj$. For comparison, the
  predictions of {\sc Disent} at LO (dashed histograms) and NLO (solid
  histograms) are included.
  \label{three}}
\end{figure}

The cross-section $(1/\sigma)(d\sigma/d(\etsbj/\etjet))$ is presented
in Fig.~\ref{two}a. The distribution of the fraction of transverse
energy contains two entries per jet and is symmetric with respect to
$\etsbj/\etjet=0.5$ by construction. The data distribution has a peak
at $\etsbj/\etjet=0.5$, which shows that the two subjets tend to have
similar transverse energies. The distribution for the difference in
pseudorapidity is shown in Fig.~\ref{two}b and also has two entries
per jet. The measured cross section has a two-peak asymmetric
structure, with a dip at $\etasbj-\etajet\sim 0$, which shows that the
subjets cannot be reconstructed too close together. 
Figure~\ref{three}a presents the normalised cross section as a
function of $|\phisbj-\phijet|$. There are two entries per jet in this
distribution. The data distribution has a peak at 
$|\phisbj-\phijet|=0.2-0.3$; the suppression at 
$|\phisbj-\phijet|\sim 0$ comes also from the fact that the subjets
cannot be resolved when they are too close together. The distribution
as a function of $\asbj$ (one entry per jet) increases as $\asbj$
increases (see Fig.~\ref{three}b). This shows that the highest
transverse energy subjet tends to be in the rear direction. This is
consistent with the asymmetric peaks observed in the $\etasbj-\etajet$
distribution.

The fixed-order QCD calculations are compared to the data in
Figs.~\ref{two} and \ref{three}. The QCD predictions give a good
description of the data in shape, within $10\%$. This shows that the
mechanism driving the subjet topology are the $q\rightarrow qg$ and
$g\rightarrow \qq$ subprocesses as implemented in the pQCD calculations.

\begin{figure}[ht]
\setlength{\unitlength}{1.0cm}
\begin{picture} (10.0,6.5)
\put (-1.0,-0.5){\epsfig{figure=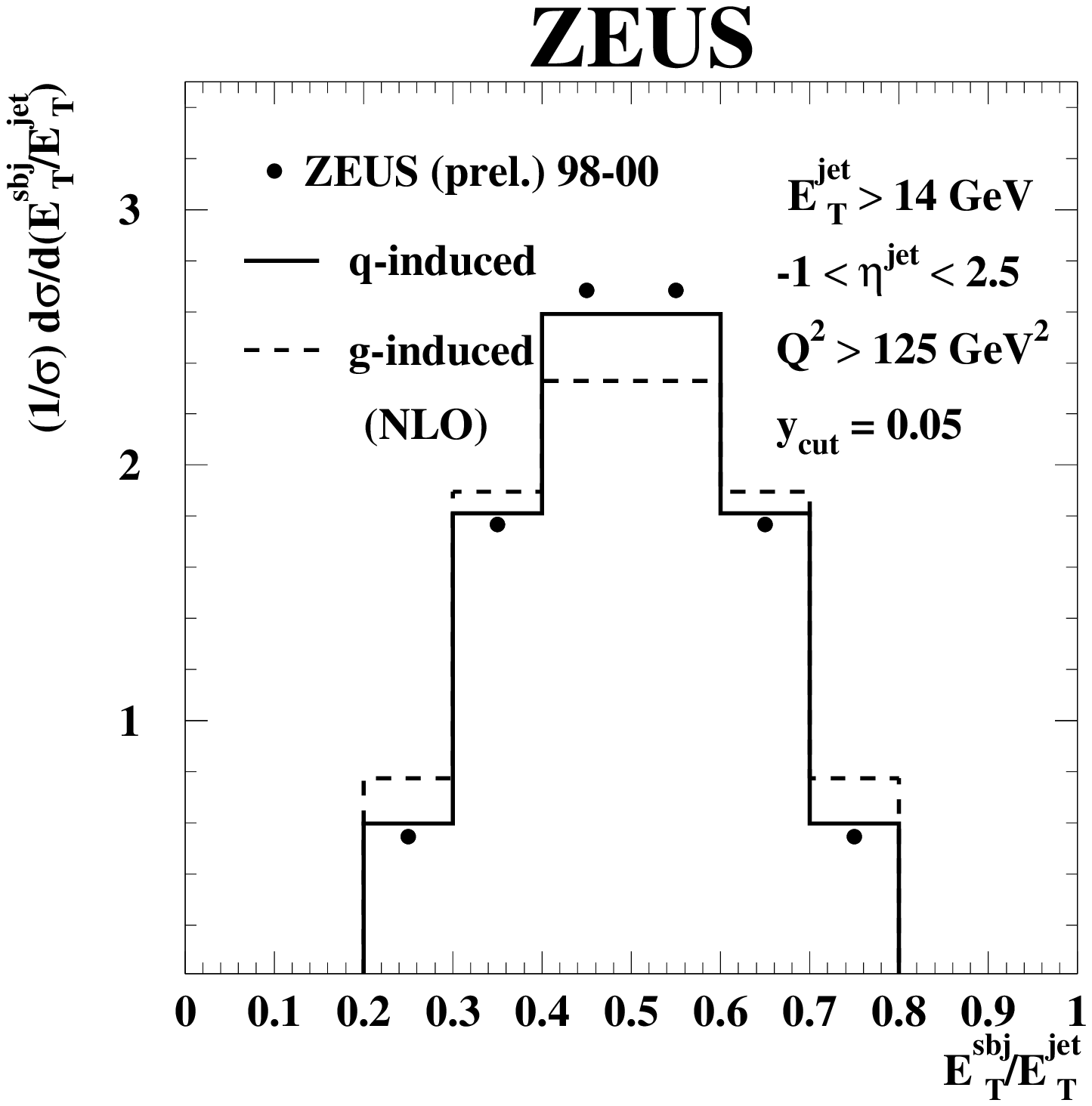,width=8.0cm}}
\put (5.0,-0.5){\epsfig{figure=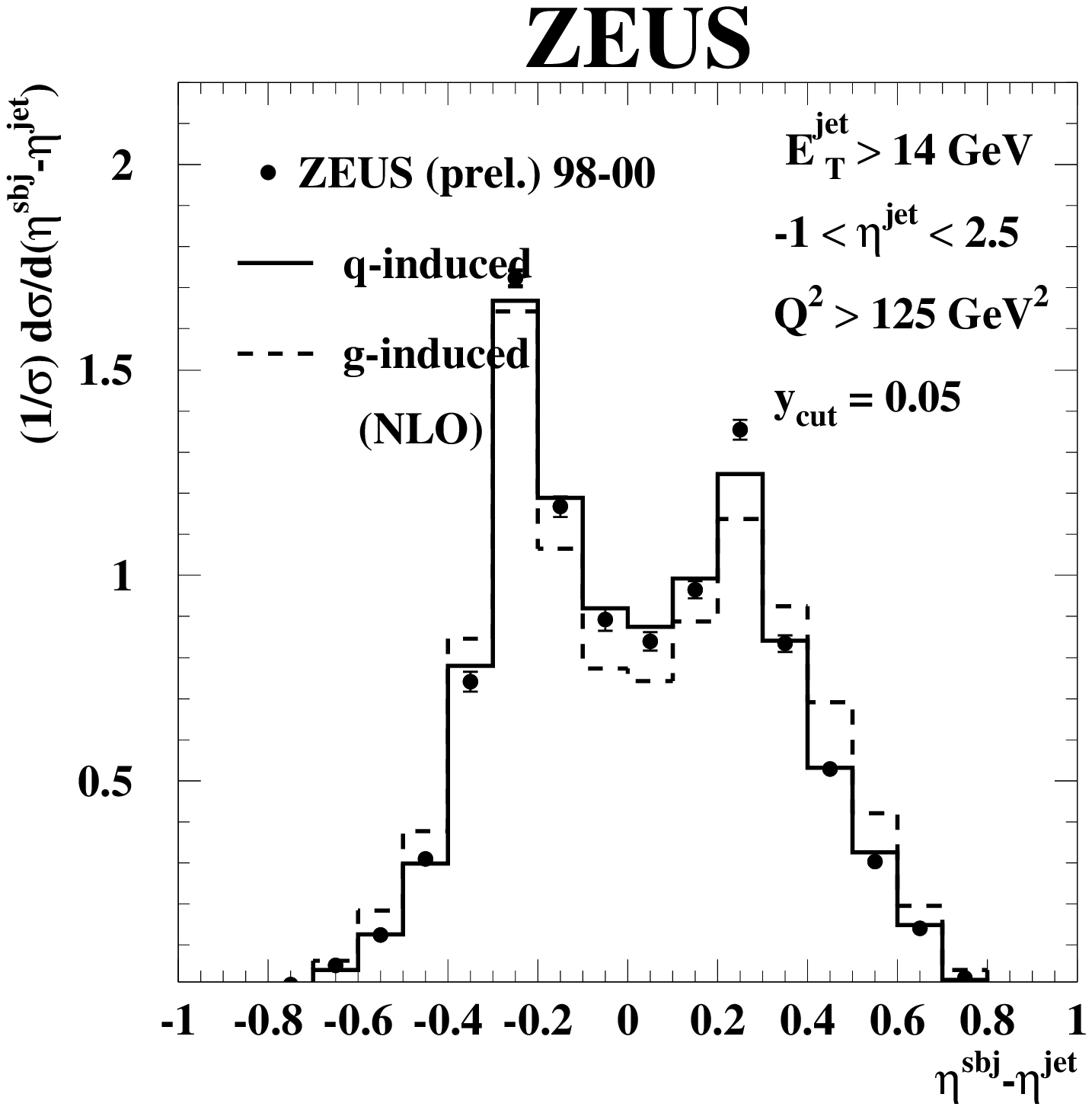,width=8.0cm}}
\put (11.0,-0.5){\epsfig{figure=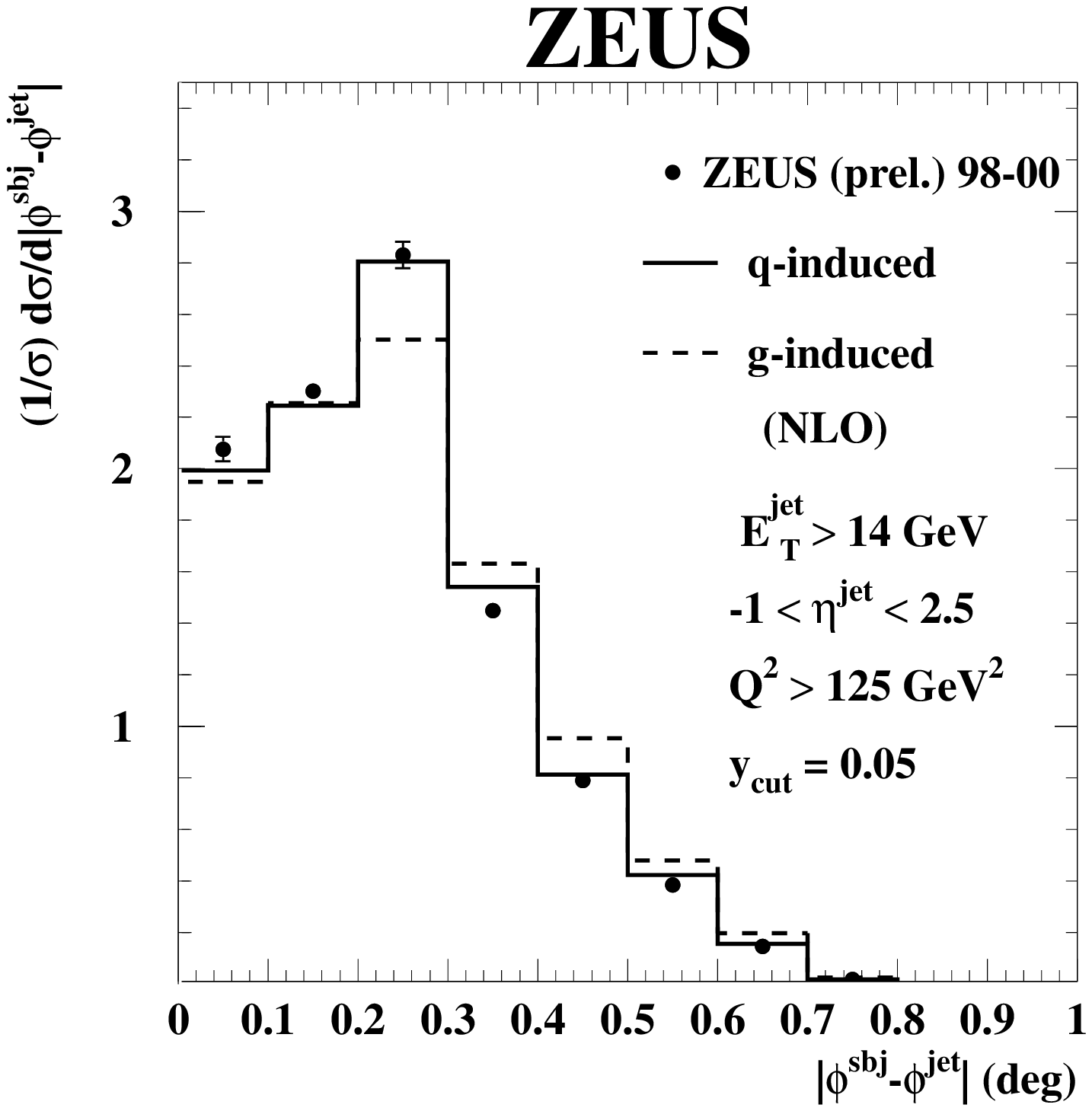,width=8.0cm}}
\put (3.0,0.0){\bf\small (a)}
\put (9.0,0.0){\bf\small (b)}
\put (15.0,0.0){\bf\small (c)}
\end{picture}
\caption{Normalised differential subjet cross sections as functions of
  (a) $\etsbj/\etjet$, (b) $\etasbj-\etajet$ and
  (c) $|\phisbj-\phijet|$. For comparison, the NLO predictions for
  quark- (solid histograms) and gluon-splitting (dashed histograms)
  are included.
  \label{four}}
\end{figure}

To study in more detail the pattern of parton radiation, the
predictions of quark- and gluon-induced processes are compared
separately with the data in Fig.~\ref{four}. The NLO calculations
predict that the two-subjet rate is dominated by quark-induced
processes: the relative contribution of quark- (gluon-) induced
processes is $82\%$ ($18\%$). The predictions for these two types of
processes are different: in quark-induced processes, the two subjets
have more similar transverse energies (Fig.~\ref{four}a) and are
closer to each other (Figs.~\ref{four}b and \ref{four}c) than in
gluon-induced processes. The comparison with the measurements shows
that the data are better described by the calculations for jets
arising from a $qg$ pair than those coming from a $\qq$ pair.

\section{Event shapes}
Event-shape observables are particularly sensitive to the details of
the non-perturbative effects of hadronisation and can be used to test
the models for these effects. In this type of analysis, the data are
compared to model predictions which combine NLO calculations and the
theoretical expectations of the power-corrections (PC) model, which is
characterised by an effective coupling $\bar\alpha_0$. The total
prediction for any event-shape observable is then given by the sum of
the perturbative and PC predictions. Previous results supported the
concept of power corrections in the approach of Dokshitzer \etal, but
a large spread of the results suggested that higher-order corrections
were needed. Now, resummed next-to-leading-logarithm (NLL)
calculations matched to NLO are available and so it is possible to
study event-shape mean values as well as distributions.

The event-shape observables studied are thrust, $T$, broadening, $B$,
the $C$ parameter and the jet mass, $M$,
$$T=\frac{\sum_i|\vec p_i\cdot \hat n|}{\sum_i|\vec p_i|} \hspace{3cm}
B=\frac{\sum_i|\vec p_i\times \hat n|}{\sum_i|\vec p_i|}$$
$$C=\frac{3\sum_{ij}|\vec p_i||\vec
  p_j|\sin^2(\theta_{ij})}{2(\sum_i|\vec p_i|)^2} \hspace{2cm}
M^2=\frac{(\sum_i E_i)^2-|\sum_i \vec p_i|^2}{(2\sum_i E_i)^2},$$
where $\hat n$ specifies the thrust or $\gamma$ axis, $\theta_{ij}$ is
the angle between two final-state particles $i$ and $j$, and $\vec
p_i\ (E_i)$ is the momentum (energy) of particle $i$.

A suitable frame in which to study event shapes at HERA is the Breit
frame since in this frame, the separation between the current jet and
the proton remnant is maximal. The event-shape variables are
reconstructed for all the particles in the current hemisphere of the
Breit frame.

\begin{figure}[ht]
\setlength{\unitlength}{1.0cm}
\begin{picture} (10.0,8.0)
\put (4.5,0.0){\epsfig{figure=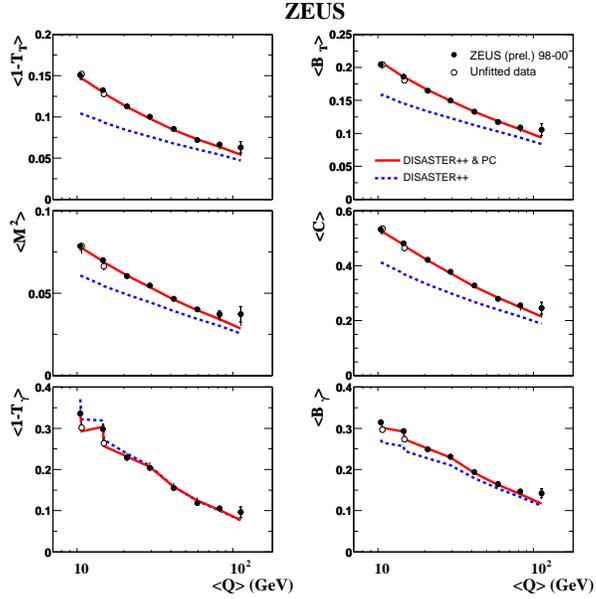,width=8.0cm}}
\end{picture}
\caption{The means of event-shape observables fitted with NLO + PC.
  \label{five}}
\end{figure}

Measurements of event-shape means have been made~\cite{eventshapes} as
a function of $Q$ for each observable in the kinematic region given by
$80<\q2<2\cdot 10^4$ \g2\ and $0.0024<x<0.6$ (see
Fig.~\ref{five}). Predictions consisting of NLO + PC calculations
have been fitted to the data, leaving $\as$ and $\bar\alpha_0$ as free
parameters. Each observable was fitted separately. The NLO predictions
were calculated using the program {\sc
  Disaster}++~\cite{disaster}. The proton PDFs have been parameterised
using the CTEQ4M~\cite{cteq} sets. A reasonable fit is obtained for
all the event-shape observables within the $\q2$ range studied.

\begin{figure}[ht]
\setlength{\unitlength}{1.0cm}
\begin{picture} (10.0,6.5)
\put (-1.0,0.0){\epsfig{figure=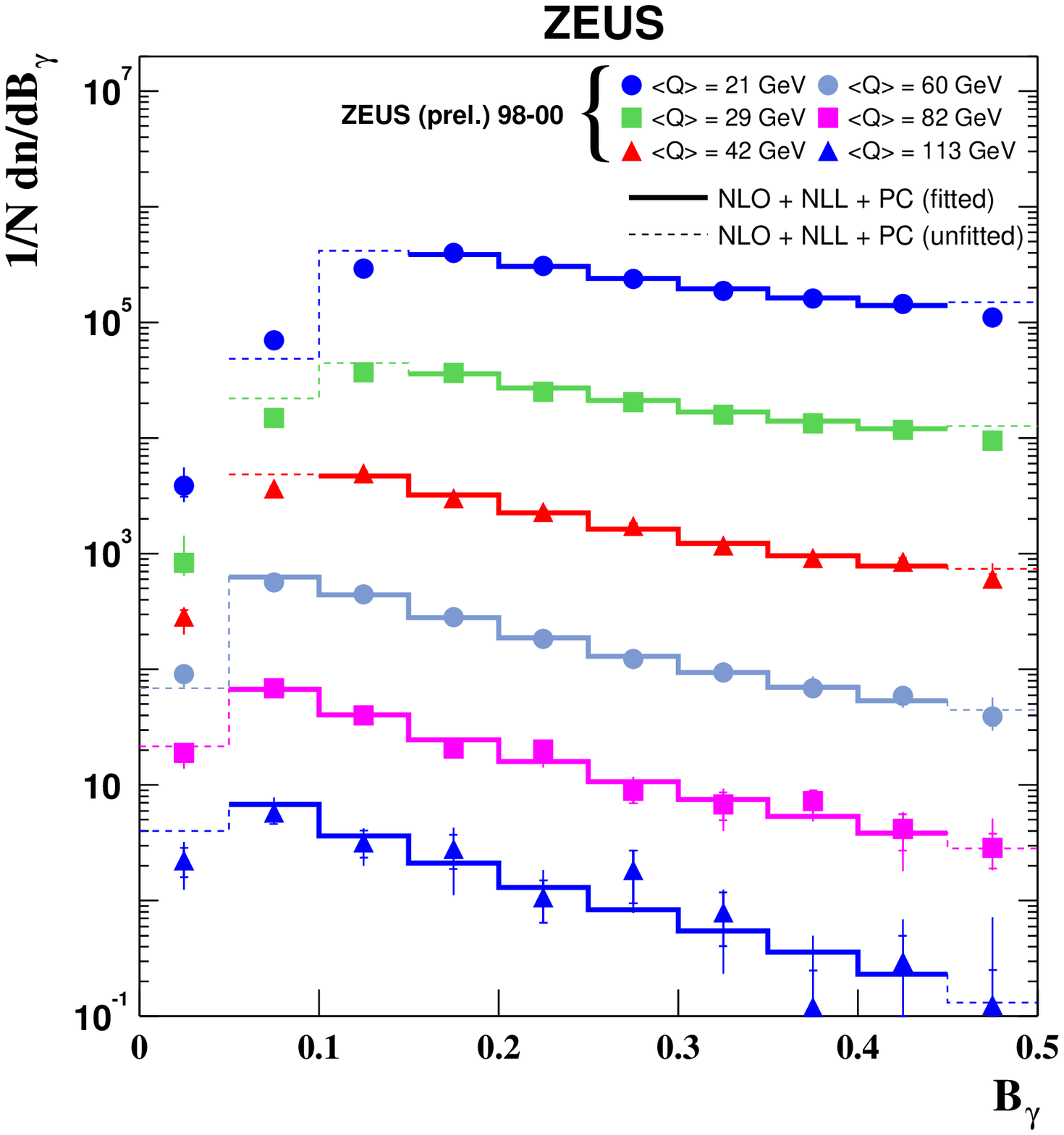,width=6.5cm}}
\put (5.0,0.0){\epsfig{figure=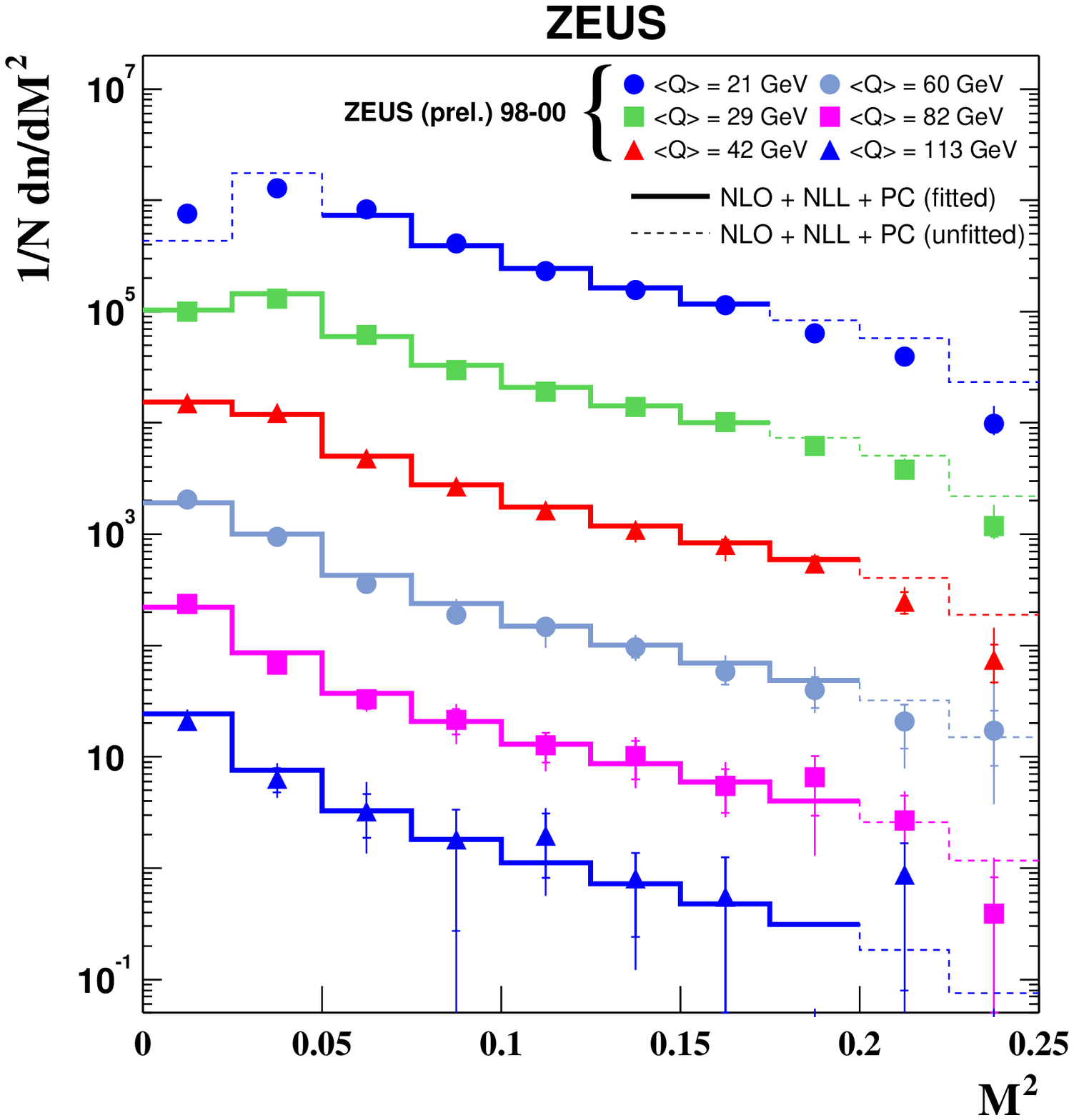,width=6.5cm}}
\put (11.0,0.0){\epsfig{figure=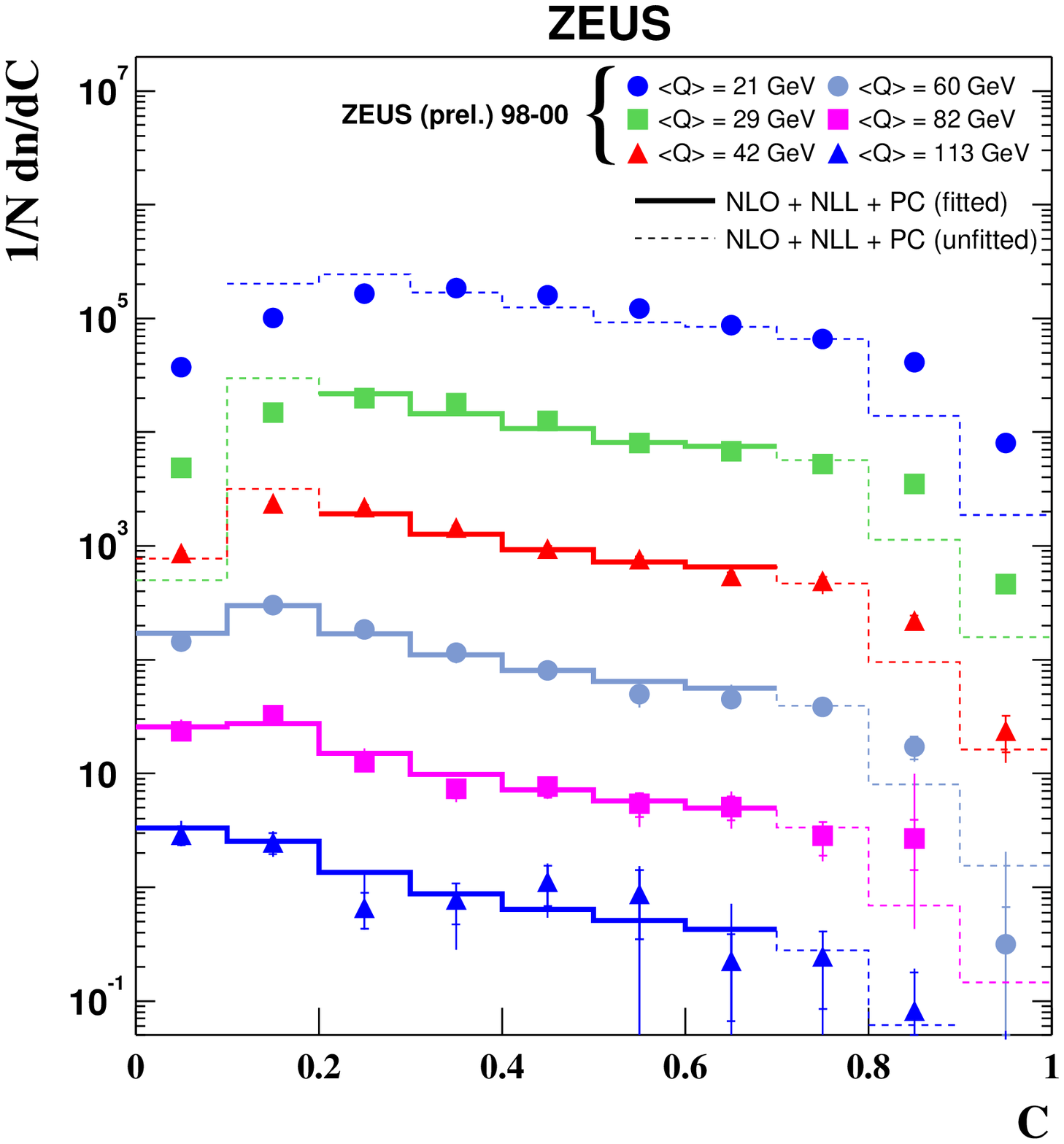,width=6.5cm}}
\put (2.5,0.0){\bf\small (a)}
\put (8.5,0.0){\bf\small (b)}
\put (14.5,0.0){\bf\small (c)}
\end{picture}
\caption{Event-shape distributions as functions of (a) $B_{\gamma}$,
  (b) $M^2$ and (c) $C$, fitted with NLL resummed calculations matched
  to NLO + PC.
  \label{six}}
\end{figure}

Figure~\ref{six} shows the event-shape differential
distributions~\cite{eventshapes} for some of the observables in
different regions of $Q$. In this case, the fit was done using NLO +
NLL + PC. The NLO predictions have been calculated using the {\sc
  Disaster}++ and {\sc Dispatch}~\cite{disp} programs with the MRST99
proton PDF sets. The PC and matched NLL predictions were calculated
using the {\sc Disresum}~\cite{disp} package. In {\sc Disresum}, the
power correction is applied as a shift of the distribution, which has
the same functional form as the power correction for the mean. For
$B_{\gamma}$, there is in addition a change in shape. The range
of the fit for each observable was restricted to the regions where the
predictions were valid. A reasonable fit is obtained for all the
event-shape observables within the restricted ranges studied.

\begin{figure}[h]
\setlength{\unitlength}{1.0cm}
\begin{picture} (10.0,15.3)
\put (0.5,7.3){\epsfig{figure=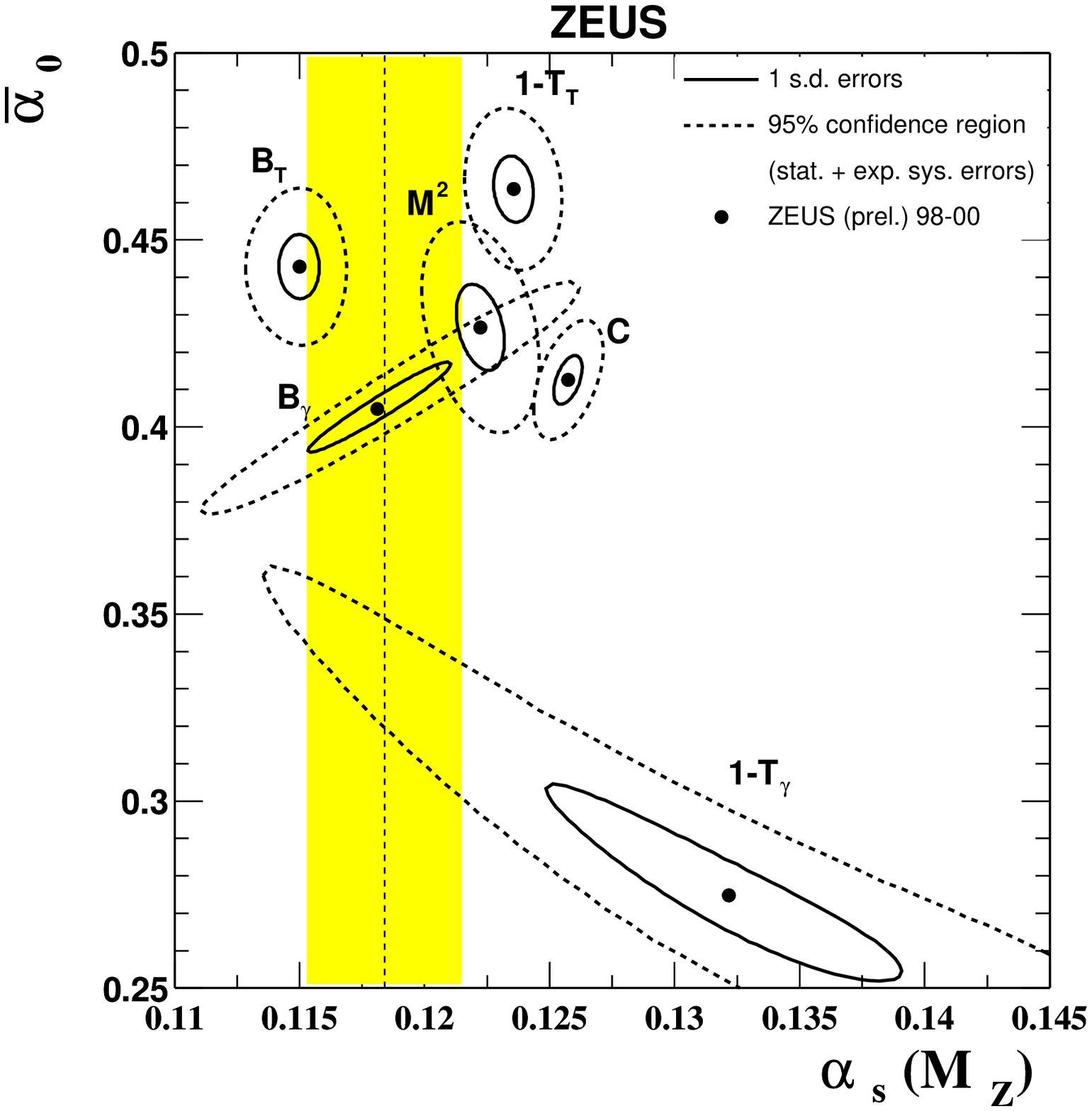,width=8.0cm}}
\put (8.5,7.3){\epsfig{figure=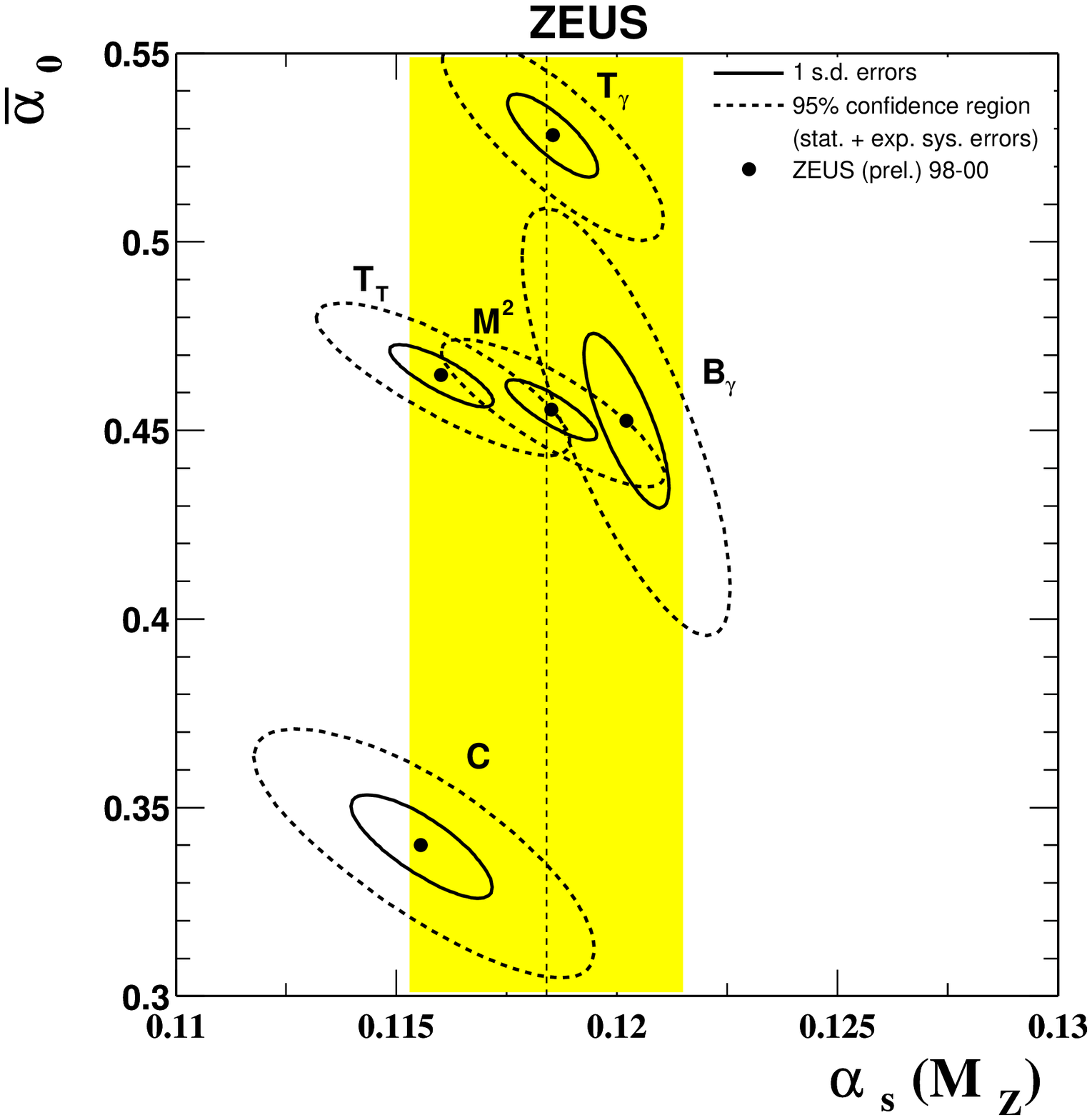,width=8.0cm}}
\put (4.5,0.0){\epsfig{figure=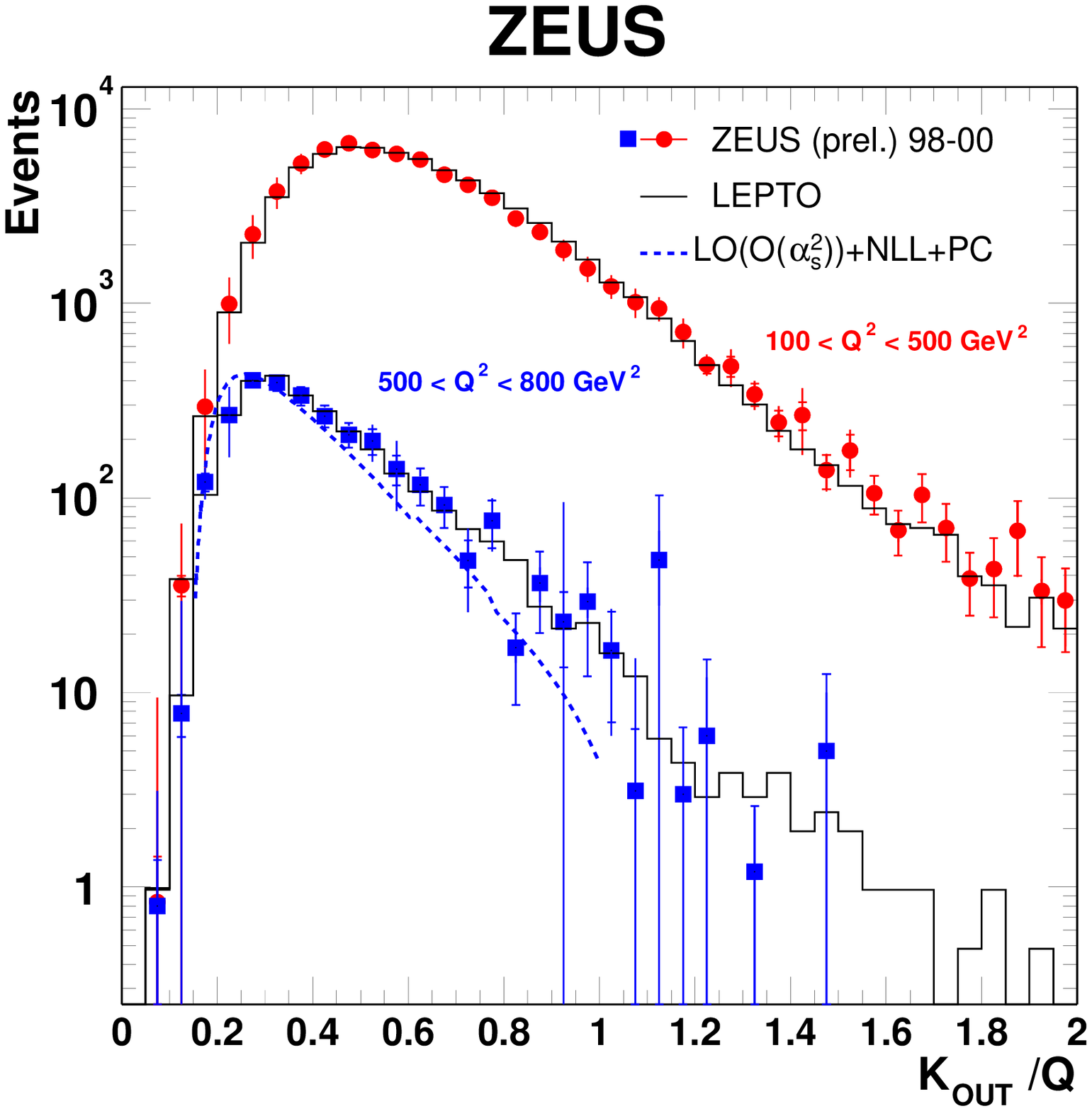,width=8.0cm}}
\put (4.5,7.5){\bf\small (a)}
\put (12.0,7.5){\bf\small (b)}
\put (8.5,0.0){\bf\small (c)}
\end{picture}
\caption{Extracted $\asz$ and $\bar\alpha_0$ parameters from (a) mean
  values and (b) distributions. (c) Distribution of $K_{\rm OUT}/Q$
  compared with {\sc Lepto} and leading-order + NLL + PC calculation.
  \label{seven}}
\end{figure}

The extracted values of $\as$ and $\bar\alpha_0$ from the means
and differential distributions are shown in Figs.~\ref{seven}a and
\ref{seven}b, respectively. It is possible to obtain a universal
value for $\bar\alpha_0$ of $0.45$ at the $10\%$ level, except for
$T_{\gamma}$ (means) and $C$ parameter (distributions). The extracted
values of $\bar\alpha_0$ and $\asz$ from the means show a dispersion
that could be due to higher-order terms. The extracted values of
$\asz$ from the distributions are consistent with the world average.

%

Measurements of an event-shape observable sensitive to higher-order
effects have been made~\cite{eventshapes}. The out-of-plane momentum,
$K_{\rm OUT}=\sum_i|{p_i^{\rm out}}|$, which is defined as the energy
flow out of the plane defined by the proton direction and the axis
which maximises the thrust, is sensitive both to perturbative and
non-perturbative contributions. The lowest non-trivial contribution
comes from non-perturbative effects or from $\oass$
contributions. Figure~\ref{seven}c shows the measurements of 
$K_{\rm OUT}/Q$ in two ranges of $\q2$. The data are well described by
the predictions of the leading-logarithm parton-shower model of 
{\sc Lepto}~\cite{lepto}. This constitutes the first comparison of
calculations which include leading-order + NLL + PC with
the data, in the high-$\q2$ range only. The description of the data by
the prediction is reasonable, but a more precise test of the model needs
higher-order calculations.

\section{Summary}
Subjet normalised cross sections have been measured in NC DIS using
$81.7$ pb$^{-1}$ of data collected with the ZEUS detector at HERA with
a centre-of-mass energy of 318 GeV. A reasonable description of the
data is obtained by the QCD predictions. This means that the pattern
of parton radiation as implemented in a NLO calculation reproduces the
behaviour of the data. In addition, the data are well described by the
calculations for jets arising from the splitting of a quark into a
quark-gluon pair.

Event-shape means and distributions have been measured in NC DIS using
$82.2$ pb$^{-1}$ of data collected with the ZEUS detector at HERA with
a centre-of-mass energy of 318 GeV. Calculations including NLO + PC,
and including resummed NLL predictions matched to NLO for the
differential distributions, give a reasonable description of the 
data. The extracted values of the power-correction parameter,
$\bar\alpha_0$, are consistent within $10\%$. The extracted values
of the strong coupling constant are consistent with the world
average. However, more theoretical input is needed to fully exploit
the potential of these measurements.

\section*{Acknowledgments}
I would like to thank my colleagues of the ZEUS Collaboration for
their help in preparing this report.


\begin{thebibliography}{99}

\bibitem{power}
Y. Dokshitzer and B. Webber, \Journal{\PLB}{352}{1997}{451}.

\bibitem{np:b406:187}
S. Catani \etal, \Journal{\NPB}{406}{1993}{187}.

\bibitem{pr:d48:3160}
S.D. Ellis and D.E. Soper, \Journal{\PRD}{48}{1993}{3160}.

\bibitem{subjets} \colab{ZEUS}, Contributed paper N-384 to the HEP2005
  International Europhysics Conference on High Energy Physics, July
  21st-27th, 2005, Lisbon, Portugal.

\bibitem{disent} S. Catani and M.H. Seymour,
  \Journal{\NPB}{485}{1997}{291}. Erratum in
  \Journal{\NPB}{510}{1998}{503}.

\bibitem{epj:c4:463}
A.D. Martin \etal, \Journal{\EPC}{4}{1998}{463};\\
A.D. Martin \etal, \Journal{\EPC}{14}{2000}{133}.

\bibitem{eventshapes} \colab{ZEUS}, Contributed paper N-381 to the HEP2005
  International Europhysics Conference on High Energy Physics, July
  21st-27th, 2005, Lisbon, Portugal.

\bibitem{disaster}
D. Graudenz, {\em Proc. of the Ringberg Workshop on New Trends in HERA
  physics}, B.A. Kniehl, G. Kr\"amer and A. Wagner (eds.). World
Scientific, Singapore (1998). Also in hep-ph/9708362 (1997);\\
D. Graudenz, Preprint hep-ph/9710244, 1997.

\bibitem{cteq}
H.L. Lai \etal, \Journal{\PRD}{55}{1997}{1280}.

\bibitem{disp}
M. Dasgupta and G.P. Salam, \Journal{\JHEP}{0208}{2002}{032}.

\bibitem{lepto}
G. Ingelman, A. Edin and J. Rathsman, \Journal{\CPC}{101}{1997}{108}.

\end{thebibliography}
\end{document}